# Search for Randall-Sundrum Graviton Excitations in the CMS Experiment

P. Traczyk[a)], G. Wrochna[a)]

*Soltan Institute for Nuclear Studies, Warsaw, Poland.*

**Abstract**

The ability of the CMS experiment to detect massive Kaluza-Klein excitations of gravitons, predicted by the Randall-Sundrum model, is studied. The search reach is estimated for the channel with the graviton, produced in proton-proton collision, decaying into a pair of electrons or muons.

a. Partially supported by the Polish Committee for Scientific Research under grant KBN 621/E-78/SPUB/P-03/DZ 5/99.

# 1 Introduction

In 1999 Arkani-Hamed, Dimopoulos and Dvali (ADD) proposed a framework for solving the gauge hierarchy problem by introducing $n$ large extra dimensions [1]. In this model, the Standard Model fields remain confined to our 4-dimensional world - new dimensions are only felt by gravity. The observed Planck scale $M_{Pl} \sim 10^{19}$ GeV is an effective scale, related to the fundamental scale $M_S$ through the volume of the compactified dimensions, $M_S^{n+2} V_n = M_{Pl}^2$. The hierarchy problem is radically solved without fine-tuning the hierarchy, but simply by removing it, setting $M_S$ equal or close to the electroweak scale $M_{EW} \sim 10^3$ GeV. Experimental consequences include possible deviations from Newtons law at small distances, and production of Kaluza-Klein excitations of the graviton in future particle colliders. Unfortunately a closer look reveals, that, while removing the original hierarchy, this model introduces a new hierarchy between the electroweak scale and the compactification scale $\mu_c = V_n^{-1/n}$, the latter ranging from around and eV to ~MeV, depending on the number of extra dimensions $n$.

Another scenario, but this time without a large compactification volume, was proposed by Randall and Sundrum (RS) [2]. Here, only one extra dimension $\phi$ is introduced, with a non-factorizable geometry based on a slice of $AdS_5$ spacetime. The extra dimension is compactified on an $S_1/Z_2$ orbifold, with fixed points $\phi = 0 , \pi$ holding two 3-branes. A solution to 5-dimensional Einstein's equations, preserving 4-dimensional Poincare invariance, is given by the metric

$$ds^2 = e^{-2kr_c\varphi}\eta_{\mu\nu}dx^{\mu}dx^{\nu} + r_c^2 d\varphi^2 \tag{1}$$

where $k \sim M_{PL}$ is the $AdS_5$ curvature, $x_\mu$ are ordinary 4-dimensional coordinates, and $r_c$ is the compactification radius. In this set-up, a fundamental 5-dimensional mass scale $m_0$, appears in a 4-dimensional theory at $\phi = \pi$, as

$$m = e^{-kr_c\pi} m_0 \tag{2}$$

The hierarchy between the Planck and electroweak scales is removed by the exponential warp factor, when $kr_c \sim$ 11-12. The compactification scale $\mu_c = 1/r_c$ is of the order of the Planck scale, and no new hierarchies appear. Because of the small compactification radius, there are no deviations from Newton's law at experimentally accessible distances. On the other hand, massive graviton excitations appear, with masses given by [3]

$$m_n = kx_n e^{-kr_c\pi} \tag{3}$$

where $x_n$ is the $n^{th}$ root of the Bessel function $J_1$. These masses are of order of a TeV, and Kaluza-Klein gravitons can be detected as massive resonances in collider experiments. Coupling of an excited graviton to matter is described by

$$L = -\frac{1}{\Lambda_\pi} T^{\mu\nu} \sum_{n=1}^{\infty} h_{\mu\nu}^{(n)} \tag{4}$$

where $T^{\mu\nu}$ is the energy-momentum tensor of the matter field, ${h_{\mu\nu}}^{(n)}$ is the $n^{th}$ excitation of the graviton, and $\Lambda_\pi$ is the scale parameter of the theory, given by

$$\Lambda_\pi = M_{Pl} e^{-kr_c\pi} \tag{5}$$

Two parameters control the properties of the RS model: the mass of the first Kaluza-Klein graviton excitation $m_G = m_1$, and the constant $c = k / M_{Pl}$, determining graviton couplings (see eq. (4) and (5)) and widths:

$$\Gamma_n = \rho m_n x_n^2 (k/M_{Pl})^2 \tag{6}$$

where $\rho$ is a constant depending on the number of open decay channels. Fig. 1, taken from [6], shows experimental and theoretical constraints no these two parameters. The $|R_5| < M_5^2$ limit is a higher-dimensional curvature bound, originally formulated in [2] as $k < M$, where $M \sim M_{Pl}$ is the fundamental, five-dimensional Planck scale. The constraint $\Lambda_\pi < 10$ TeV assures no new hierarchy appears between $m_{EW}$ and $\Lambda_\pi$. The remaining two curves represent experimental bounds, obtained from lepton and jet pair production analysis at the Tevatron, and from a global fit to the electroweak oblique parameters $S$ and $T$. The allowed region, enclosed by the four curves, can, according to [6] be completely covered by the LHC. This, however, is based on an analysis assuming an ideal detector. The purpose of this paper will be to obtain the search reach for the CMS experiment and verify the above statement taking into account detector effects.

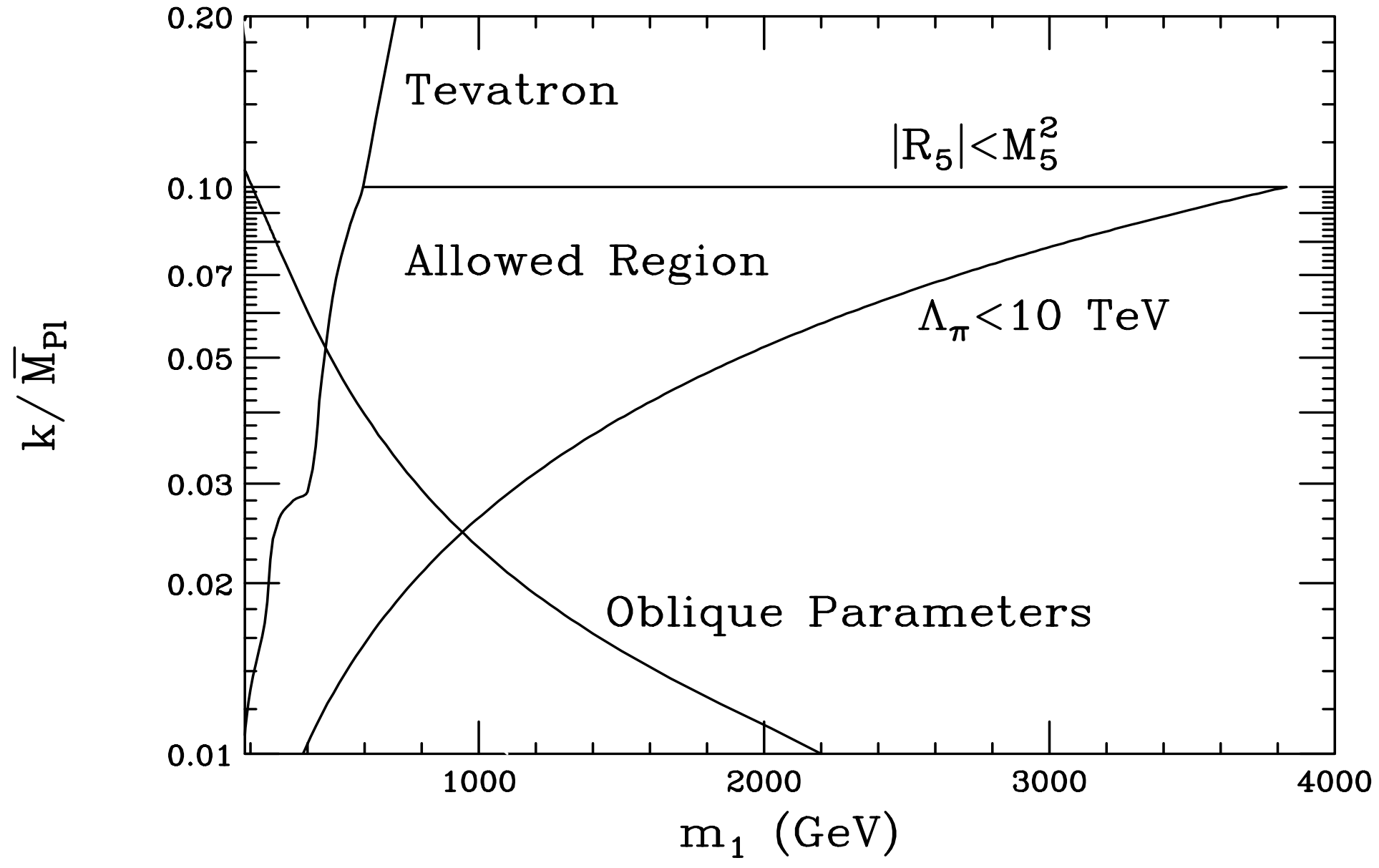

**Figure 1:** Experimental and theoretical constraints on the RS model parameters $c = k / M_{Pl}$ and $m_G = m_1$ [6].

# 2 Monte Carlo simulations

## 2.1. Signal and background simulation

The process chosen to be the "signal from the extra dimension" is the production of the first excited KK graviton state, and it's decay into a pair of leptons[a)]. A clear signature of this process are high-$p_t$ leptons, easy to identify and trigger. The Standard Model background in this channel consists mostly of the Drell-Yan process $pp \to Z^0/\gamma \to ll$. Simulations were done independently with two event generators, PYTHIA 6.157 [4] and HERWIG 6.201 [5]. Both have production and decay of RS graviton resonances implemented in the original code.

The Monte Carlo generators were set to produce gravitons in proton-proton collisions at 14 TeV centre-of-mass energy. Leptons were identified among the final state particles, and an initial detector acceptance cut ($|\eta| < 2.4$ for muons, and $|\eta| < 3.0$ for electrons) was applied. Simulations were done for 20 sets of model parameters, covering the range suggested in [6].

Fig. 2 shows a comparison between results from the two event generators. The numbers of events shown correspond to one year of LHC running at the design luminosity of $10^{34}$ cm$^{-2}$s$^{-1}$, giving an integrated luminosity of 100 fb$^{-1}$. The graviton can be seen as a peak in the invariant mass distribution of leptons. The graviton decays into all SM particles, with leptonic decays being only about 4% of all decays. Low energy secondary leptons, coming from tau, W and Z decays, are responsible for most of the peak's asymmetry. Dotted histogram shows leptons coming directly from graviton decay (for data from Herwig).

Results from Pythia and Herwig are not identical, the shape and height of the resonance are different in the two cases. The origin of these differences is show in Fig. 3, showing the mass distribution of generated gravitons. The graviton generated with Herwig is symmetric, whereas the shape generated by Pythia resembles more the one seen in Fig. 2. The reason is simple: Herwig calculates graviton production cross section in the maximum, and smears the particle's mass according to the Breit-Wigner distribution, while Pythia works the other way around - the cross section is calculated as a function of the graviton mass. In proton-proton collisions, the probability of finding a pair of partons with a given centre-of-mass energy is a rapidly falling function of that energy, hence the asymmetry. The approximation used in Herwig is good for narrow resonances, but not in the case of a 20 GeV-wide graviton. Calculated cross sections are shown in Tab. 1. Background simulation results from both generators are consistent with each other, as shown in Fig. 4.

a. Throughout this paper the term "lepton" refers to electrons and muons only.

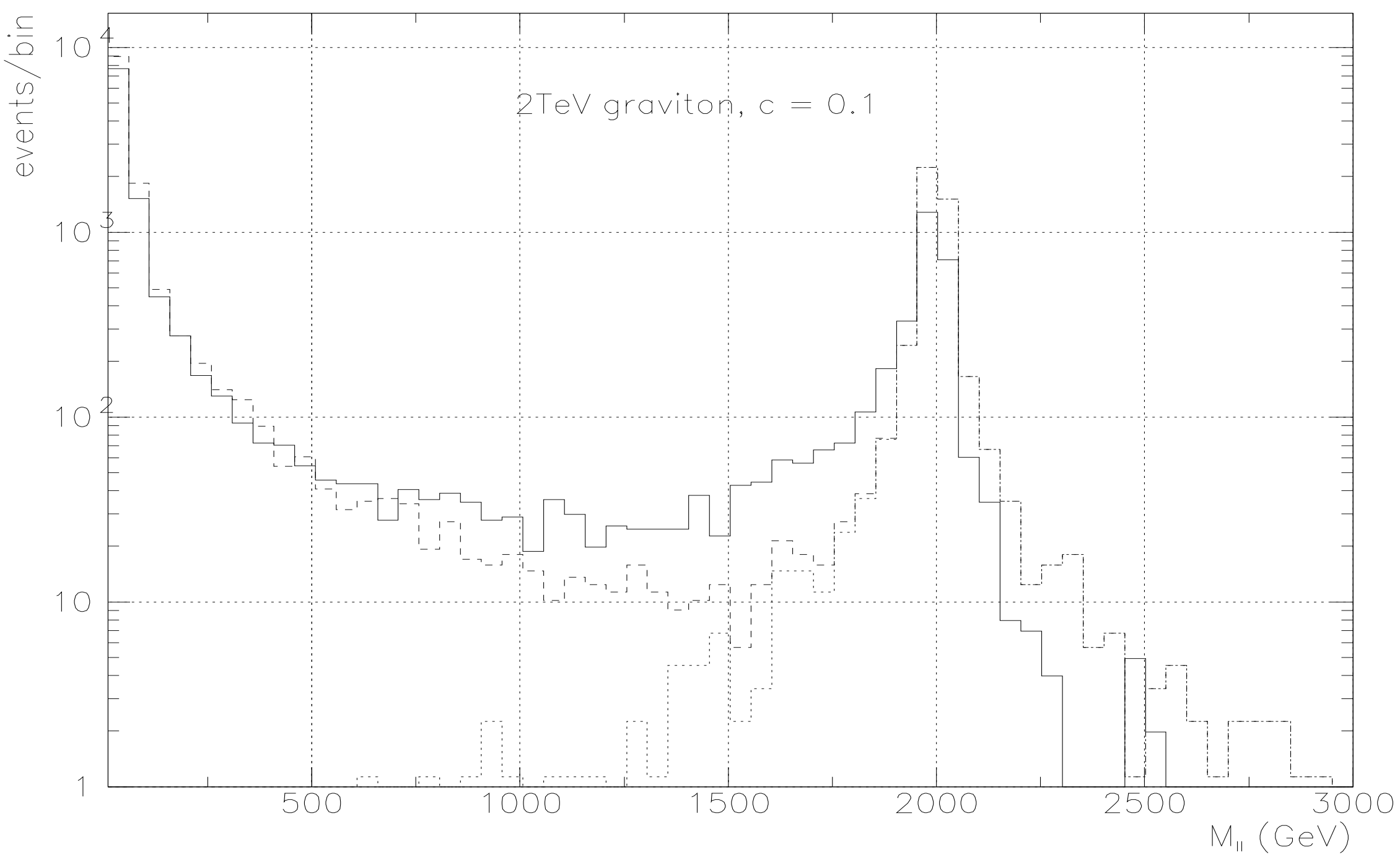

**Figure 2:** Invariant mass distribution of lepton pairs from a 2 TeV graviton simulation (with *c* = 0.1). Solid line shows results from Pythia, dashed line from Herwig. Dotted histogram shows what the distribution would look like if only leptons coming directly from graviton decay were taken into account (for Herwig data).

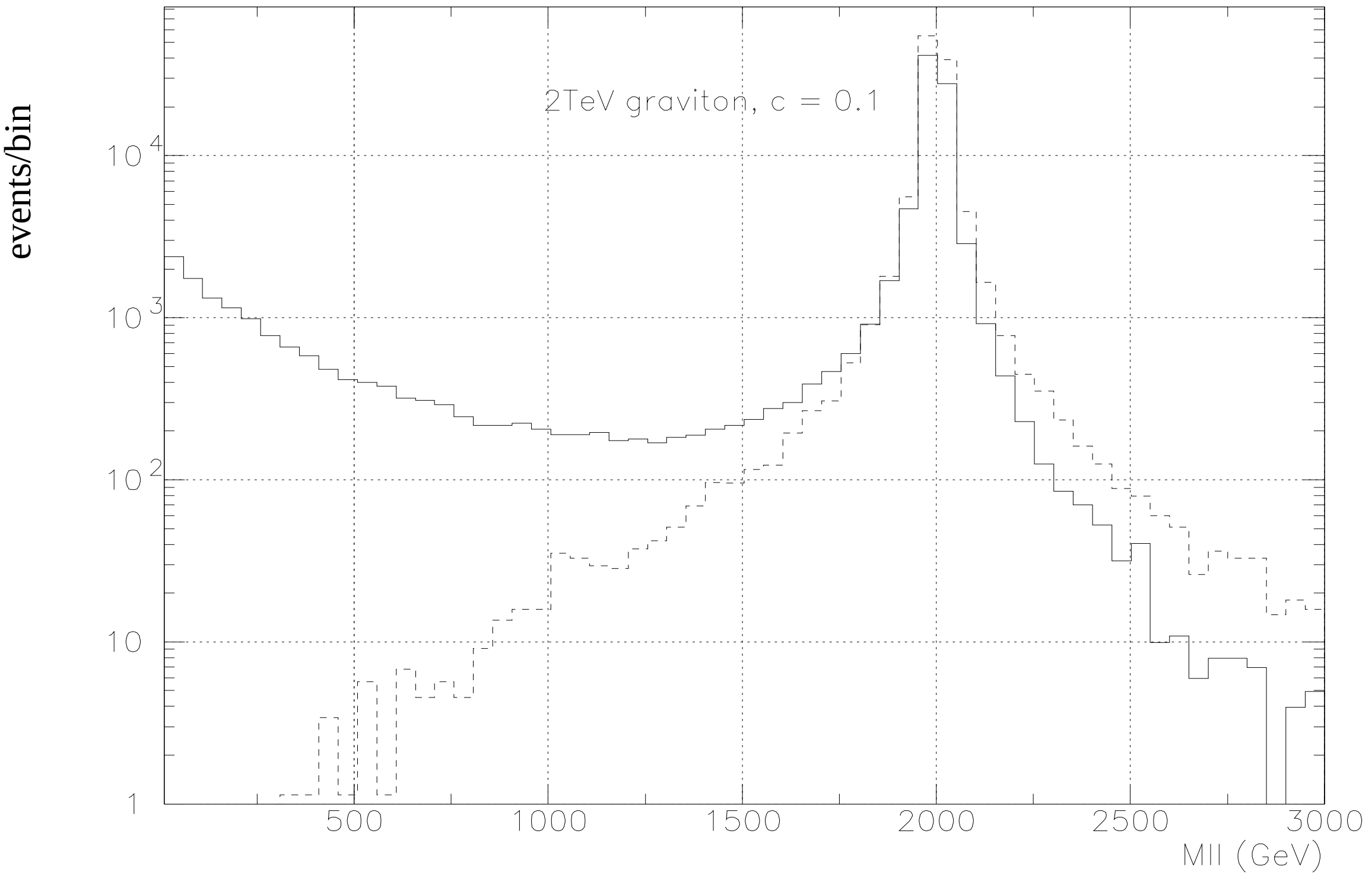

**Figure 3:** Distribution of graviton mass for Pythia (solid line), and Herwig (dashed line), normalized for 100 $fb^{-1}$ of integrated luminosity.

**Table 1:** Cross sections (in mb) for the simulated graviton production and decay process.

| Generator | $m_G$ [GeV] | *c* 0.01 | 0.02 | 0.05 | 0.07 | 0.10 | 0.20 |
|---|---|---|---|---|---|---|---|
| PYTHIA | 1000 | $3.93 \times 10^{-10}$ | $1.58 \times 10^{-09}$ | $1.00 \times 10^{-08}$ | | $4.14 \times 10^{-08}$ | |
| | 1500 | $4.53 \times 10^{-11}$ | | $1.16 \times 10^{-09}$ | | $4.99 \times 10^{-09}$ | |
| | 2000 | $8.44 \times 10^{-12}$ | | $2.20 \times 10^{-10}$ | | $9.90 \times 10^{-10}$ | |
| | 3000 | $5.69 \times 10^{-13}$ | | $1.59 \times 10^{-11}$ | $3.48 \times 10^{-11}$ | $8.57 \times 10^{-11}$ | |
| | 4000 | | | $1.95 \times 10^{-12}$ | | $1.46 \times 10^{-11}$ | $1.67 \times 10^{-10}$ |
| | 5000 | | | | | $4.27 \times 10^{-12}$ | $6.09 \times 10^{-11}$ |
| | 6000 | | | | | | $2.80 \times 10^{-11}$ |
| HERIWG | 1000 | $5.77 \times 10^{-10}$ | $2.31 \times 10^{-09}$ | $1.44 \times 10^{-08}$ | | $5.73 \times 10^{-08}$ | |
| | 1500 | $6.50 \times 10^{-11}$ | | $1.62 \times 10^{-09}$ | | $6.44 \times 10^{-09}$ | |
| | 2000 | $1.14 \times 10^{-11}$ | | $2.86 \times 10^{-10}$ | | $1.13 \times 10^{-09}$ | |
| | 3000 | $6.84 \times 10^{-13}$ | | $1.71 \times 10^{-11}$ | $3.34 \times 10^{-11}$ | $6.78 \times 10^{-11}$ | |
| | 4000 | | | $1.56 \times 10^{-12}$ | | $6.23 \times 10^{-12}$ | $2.48 \times 10^{-11}$ |
| | 5000 | | | | | $6.75 \times 10^{-13}$ | $2.84 \times 10^{-12}$ |
| | 6000 | | | | | | $3.56 \times 10^{-13}$ |

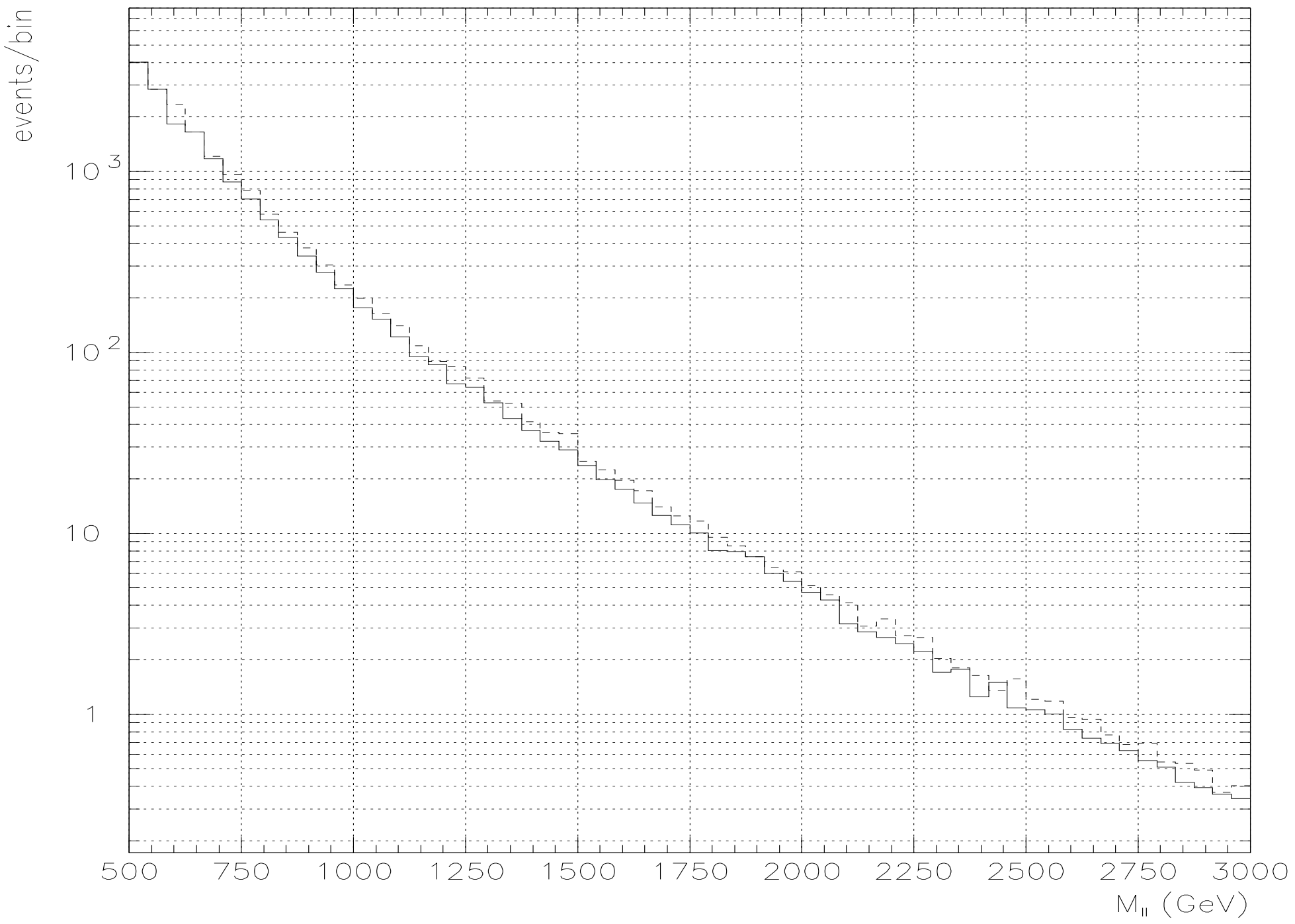

**Figure 4:** Background simulation results. Invariant mass distribution of lepton pairs from Pythia (solid line) and Herwig (dashed line).

## 2.2. Detector simulation

Detector response was simulated by passing generated signal and background leptons through CMSJET 4.703 program [7]. Fig. 5 shows the impact of detector effects on the signal. Some of the signal is lost, and the peak is smeared, especially in the case of muons. This is natural, since muon momentum measurement is based on the particle track's curvature in magnetic field, momentum resolution of the muon system can be approximated by ([8])

$$\frac{\Delta p}{p}\% = 4\%\sqrt{p} \tag{7}$$

where the muon momentum $p$ is measured in TeV. Electron energy is measured in a calorimeter, with resolution parametrized by

$$\frac{\Delta E}{E} = \frac{a}{\sqrt{E}} \oplus \frac{b}{E} \oplus c \tag{8}$$

where $E$ is measured in GeV. The $a$, $b$, and $c$ parameter values used in CMSJET are of order of, respectively, 5%, 20% and 0,5%.

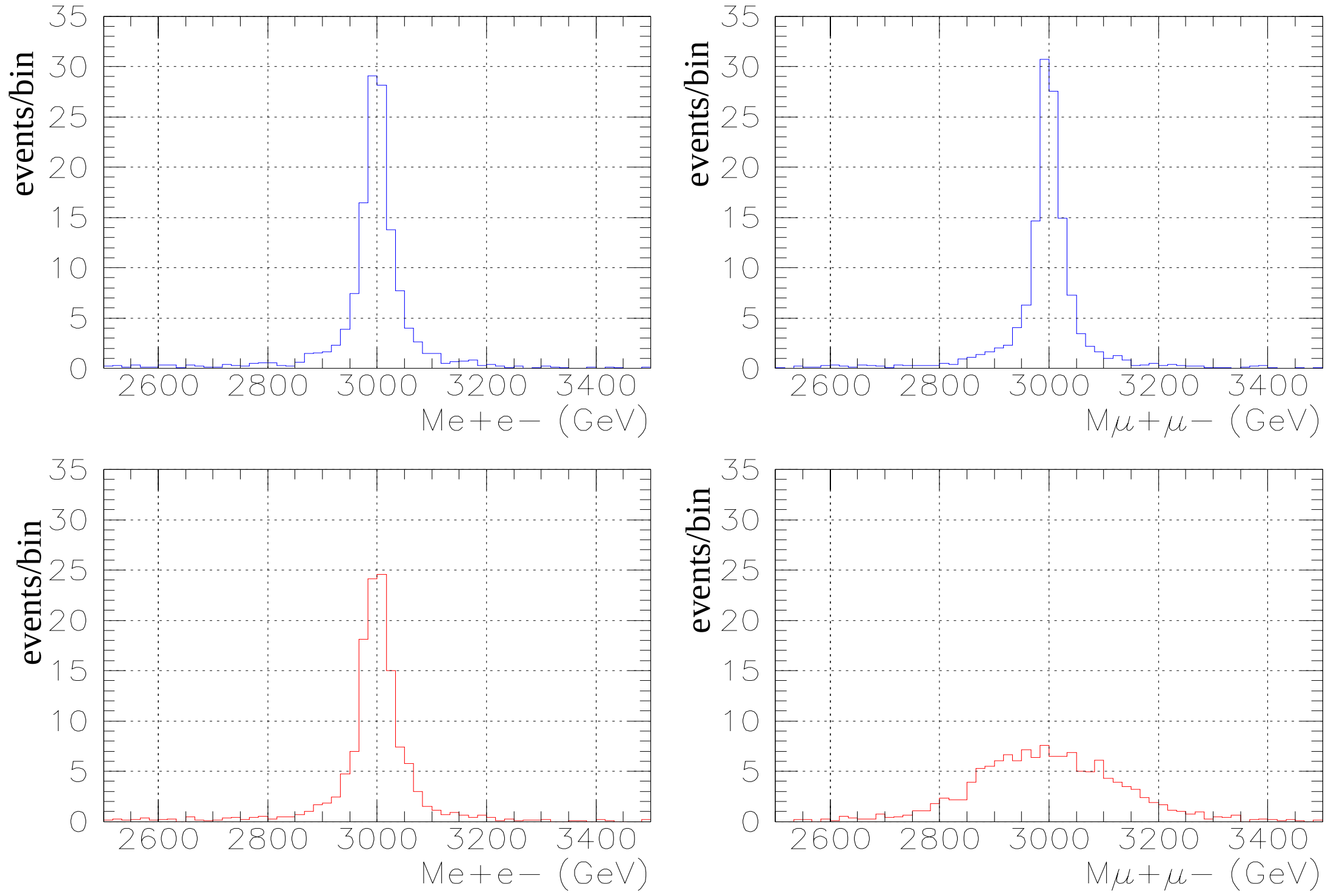

**Figure 5:** Electron and muon pair invariant mass distributions for $m_G$ = 3 TeV and $c$ = 0.1. The upper two plots show results directly from the Herwig generator, the lower two show data after detector simulation.

# 3 Data analysis

For every set of model parameters an invariant mass window around the graviton mass was chosen, and the number of signal and background events inside was calculated. Muons and electrons were treated independently, because of the difference in observed resonance widths induced by the detector. Tab. 2 shows the graviton widths, as calculated by Herwig. The total observed width can be approximated by

$$\Gamma_{obs} \approx \sqrt{\Gamma_1^2 + 2(\Delta E)^2} \tag{9}$$

where $\Gamma_1$ is the graviton width, and $\Delta E$ is the electron or muon energy measurement uncertainty (equations (7) and (8)) for a particle with momentum $m_G$.

**Table 2:** Widths of simulated gravitons.

| $m_G$ [GeV] | | | $c$ | | | |
|---|---|---|---|---|---|---|
| | **0.01** | **0.02** | **0.05** | **0.07** | **0.10** | **0.20** |
| 1000 | 0.1 | 0.6 | 3.5 | | 14.1 | |
| 1500 | 0.2 | | 5.3 | | 21.3 | |
| 2000 | 0.3 | | 7.1 | | 28.4 | |
| 3000 | 0.4 | | 10.6 | 20.9 | 42.6 | |
| 4000 | | | 14.2 | | 56.8 | 227.2 |
| 5000 | | | | | 71.0 | 284.0 |
| 6000 | | | | | | 340.8 |

Width of the mass window was set to $\pm 2\Gamma_{obs}$. Expected number of signal and background events ($N_S$ and $N_B$) inside the window was calculated for 100 fb$^{-1}$ integrated luminosity. The number of actually observed background leptons fluctuates according to the Poisson distribution, with a dispersion of $\pm\sqrt{N_B}$. The total number of signal and background events $N^{obs} = N_S^{obs} + N_B^{obs}$ is also described by the Poisson distribution. The probability $P$, that it will be less than $N_B + 2\sqrt{N_B}$ is given by:

$$P(N^{obs} < N_B + 2\sqrt{N_B}) = \sum_{k=0}^{N_B + 2\sqrt{N_B}} \frac{(N_S + N_B)^k}{k!} e^{-(N_S + N_B)} \qquad (10)$$

and when $1 - P = 0.05$, the hypothesis can be excluded at 95% confidence level when no signal is observed.

The figures below show examples of signal and background distributions (from Pythia) for a few sets of model parameters. Numbers of events from both generators are shown in Tab. 3 and Tab. 4. There is a significant difference in electron and muon signal in the data from Pythia. This effect arises from the difference in window widths and from the asymmetric shape of the graviton resonance. The muon window is wider, and more particles from the peak's left slope fall into the window. In the data from Herwig, where the resonance is symmetric, no such effect is observed.

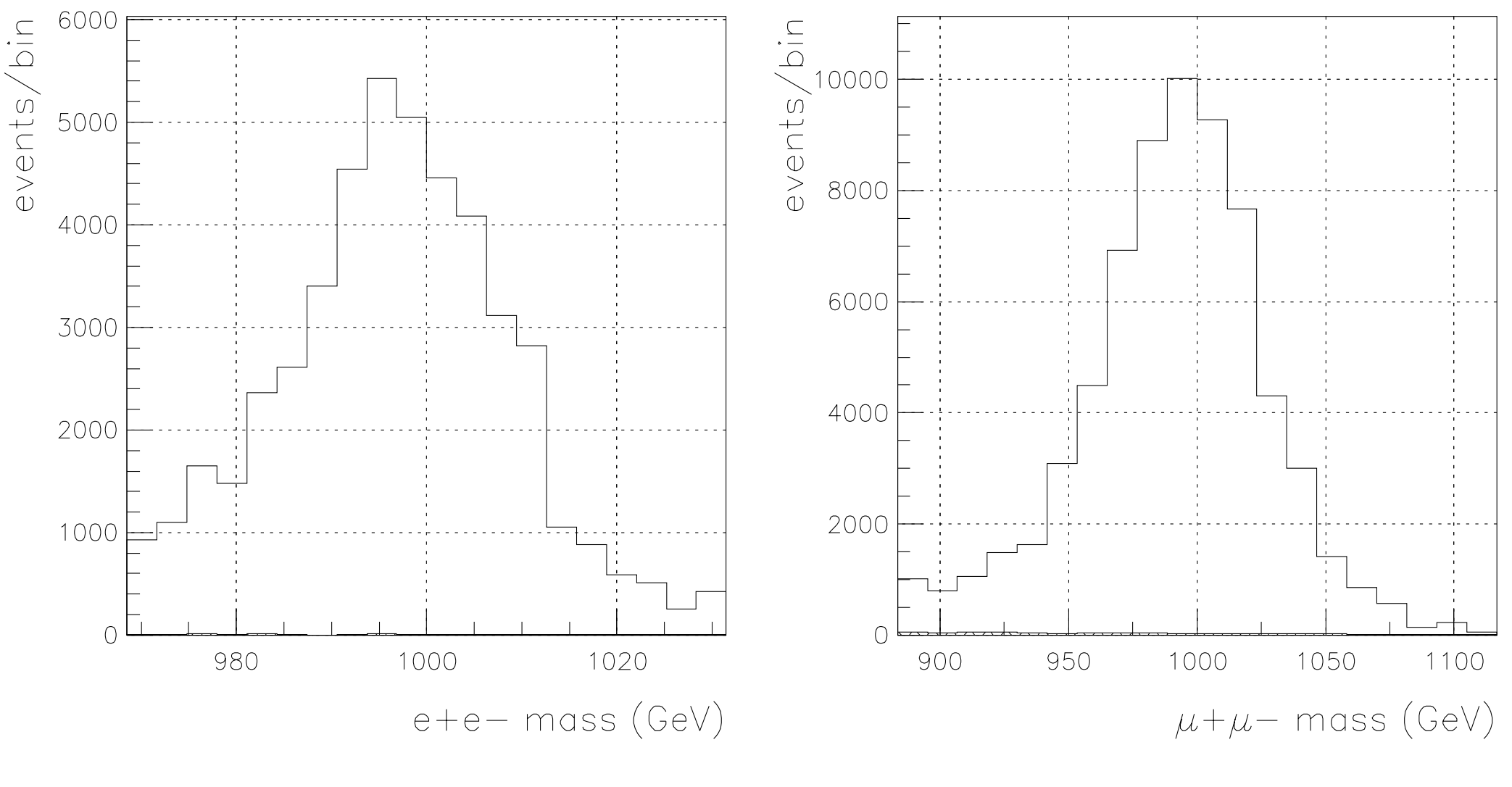

**Figure 6:** Signal distribution for $m_G$ = 1000 GeV and $c$ = 0.1. Background, shown with the hatched histogram, is practically invisible.

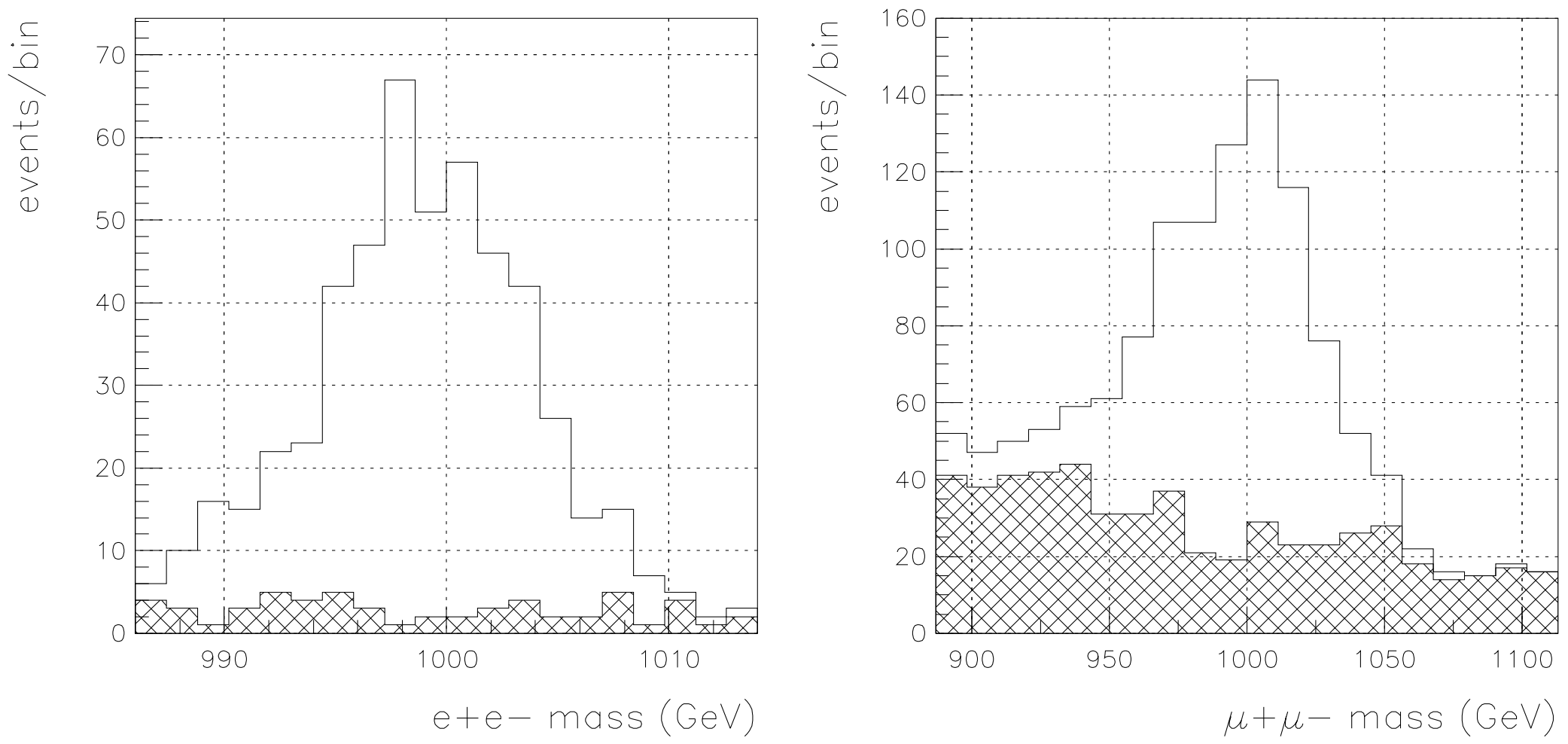

**Figure 7:** Signal and background distribution for $m_G$ = 1000 GeV and $c$ = 0.01.

The most optimistic case is shown in Fig. 6. A light, strongly coupled graviton is impossible to miss. Background is negligible, totally overwhelmed by the signal. Event numbers drop by a factor of 10 after reducing the coupling ten times (Fig. 7). Fig. 8 shows a more difficult case, with graviton mass of 2 TeV, and $c$ = 0.01. This is a rare particle, with expected 11 signal electron events over 3 events in background, giving a 3.3% chance of accidentally excluding the model. The muon background is higher, due to a wider mass window, and with 15 signal and 44 background events the probability of loosing the signal is 48%. Increasing the coupling to a value of 0.2 allows a 4 TeV graviton to be observed (Fig. 10). Increasing the mass further causes the signal to drop, the heaviest case is shown in Fig. 11. A 5 TeV graviton is practically out of reach at this luminosity.

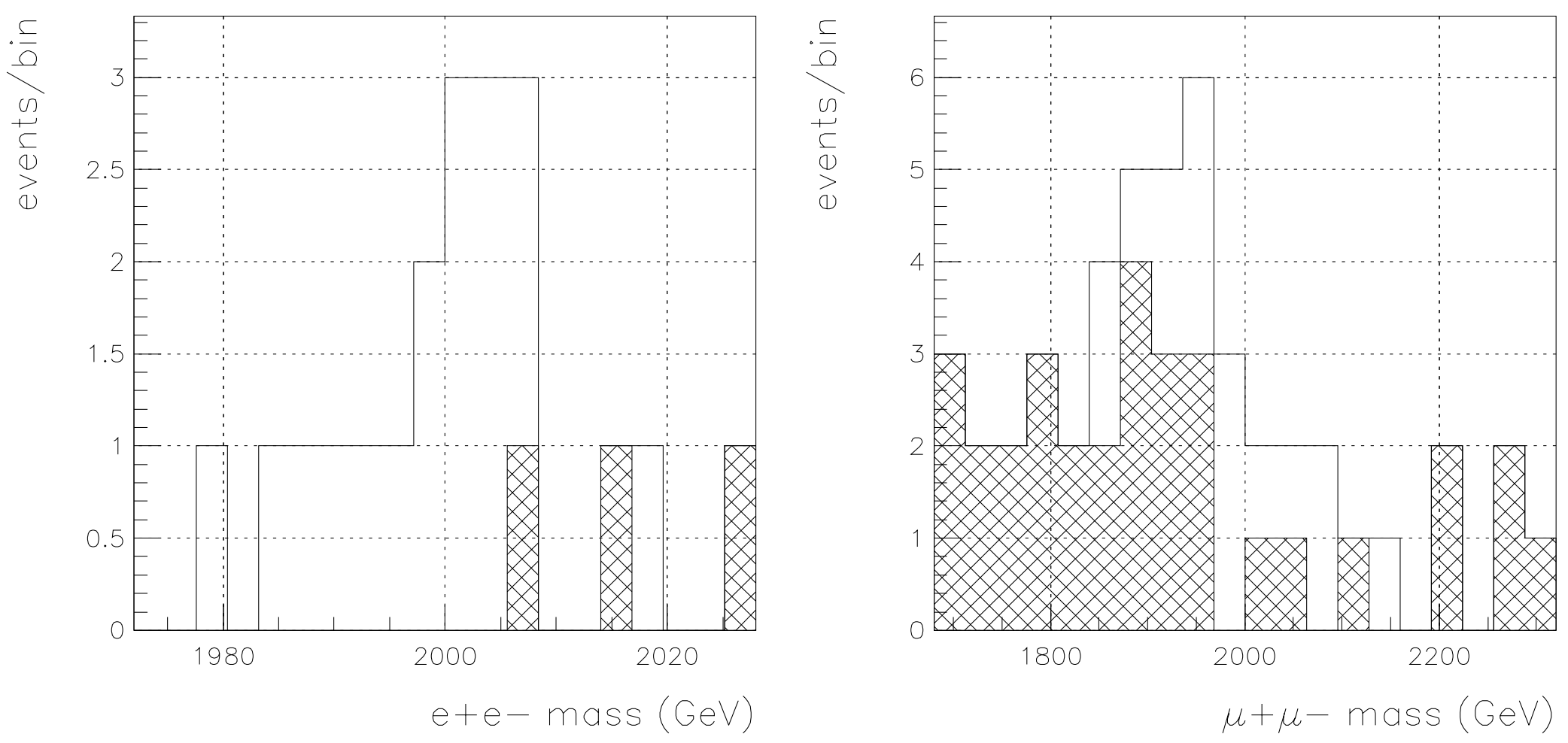

**Figure 8:** Signal and background distribution for $m_G$ = 2000 GeV and $c$ = 0.01.

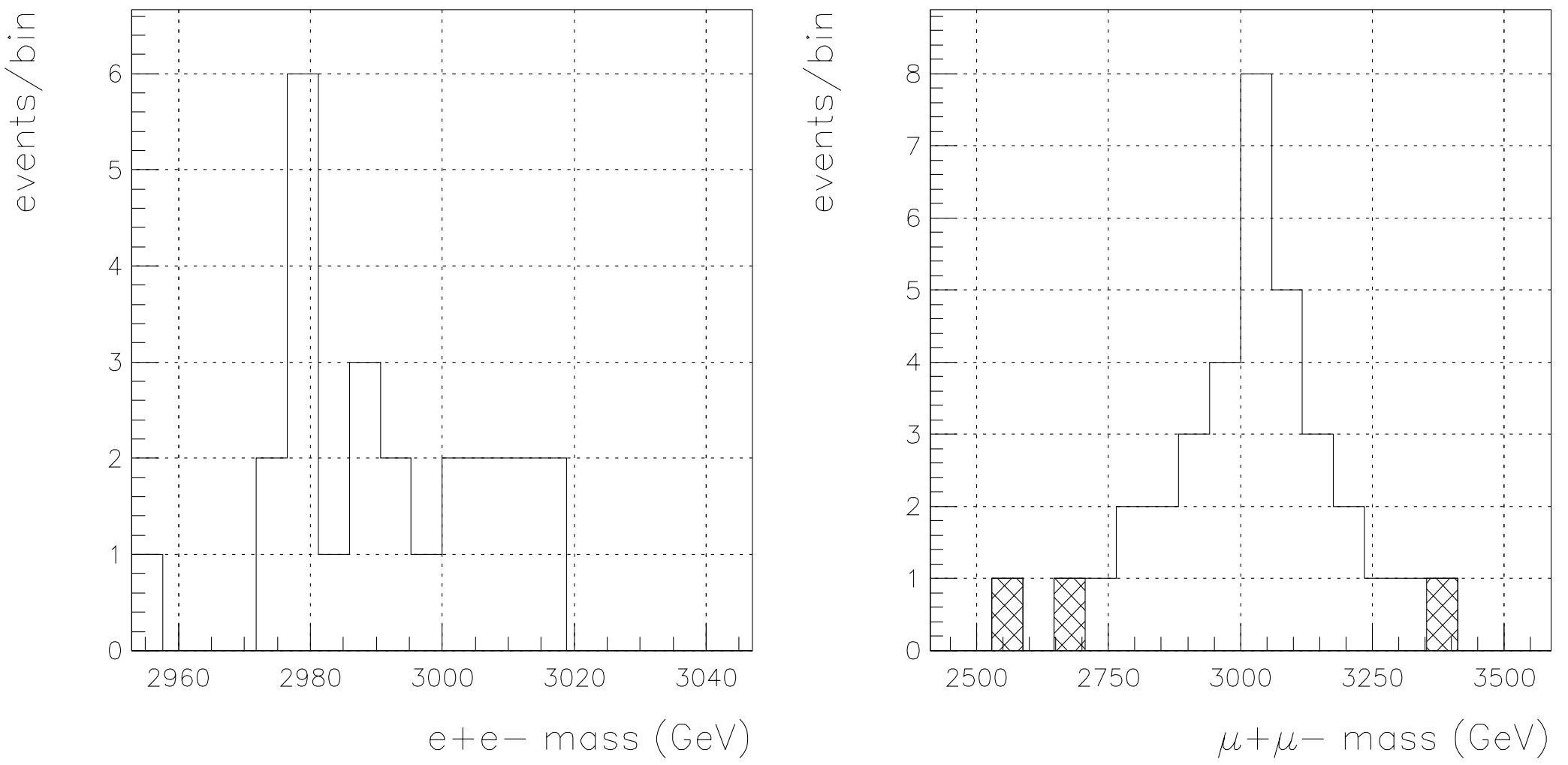

**Figure 9:** Signal and background distribution for $m_G$ = 3000 GeV and $c$ = 0.05.

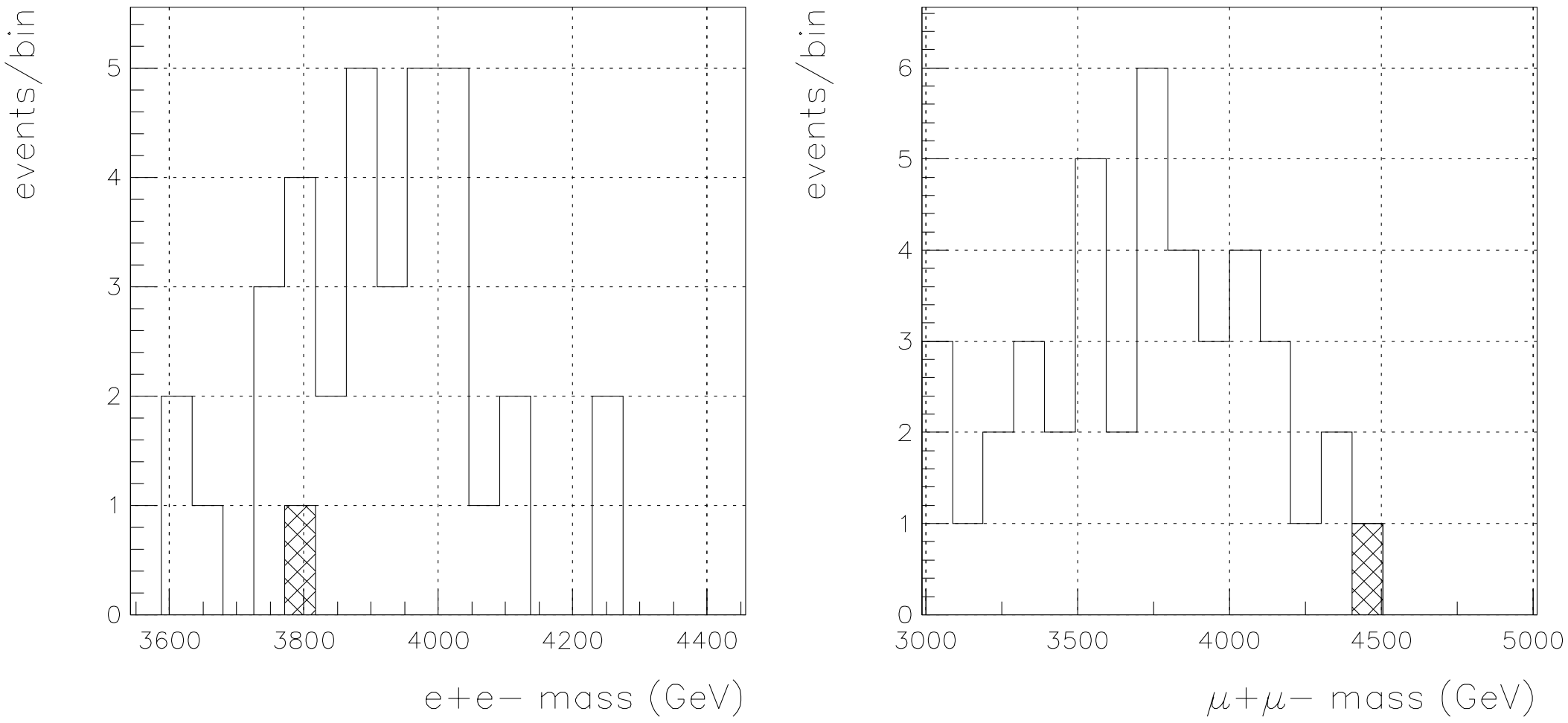

**Figure 10:** Signal and background distribution for $m_G$ = 4000 GeV and $c$ = 0.2.

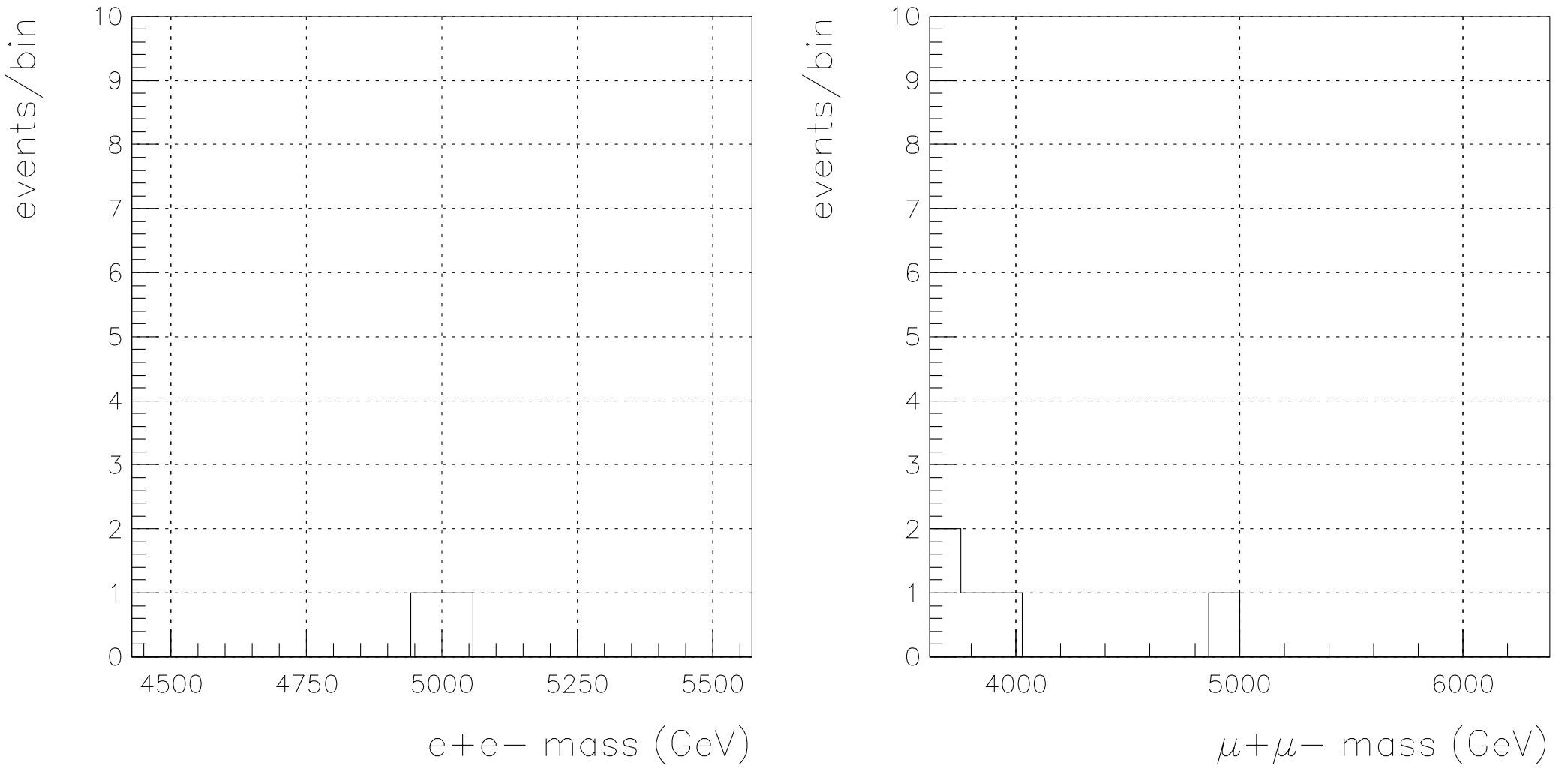

**Figure 11:** Signal and background distribution for $m_G$ = 5000 GeV and $c$ = 0.2.

**Table 3:** Signal/background events from Pythia.

| Signal | $m_G$ [GeV] | 0.01 | 0.02 | $c$ 0.05 | 0.07 | 0.10 | 0.20 |
|---|---|---|---|---|---|---|---|
| $e^+e^-$ | 1000 | **490**/65 | **2028**/65 | **11770**/72 | | **45953**/148 | |
| | 1500 | **56**/13.8 | | **1361**/15.4 | | **5244**/30 | |
| | 2000 | **10.8**/3.1 | | **255**/3.6 | | **1042**/7.2 | |
| | 3000 | **0.7**/0.3 | | **15.4**/0.4 | **31**/0.4 | **69**/0.7 | |
| | 4000 | | | **1.5**/0.04 | | **6.7**/0.1 | **33**/0.3 |
| | 5000 | | | | | **0.8**/0.01 | **3.2**/0.04 |
| | 6000 | | | | | | **0.3**/0.01 |
| $\mu^+\mu^-$ | 1000 | **716**/540 | **2861**/540 | **16120**/541 | | **65370**/557 | |
| | 1500 | **80**/138 | | **2039**/139 | | **8234**/141 | |
| | 2000 | **15**/44 | | **379**/44 | | **1477**/45 | |
| | 3000 | **1.1**/6.2 | | **25**/6.2 | **50**/6.2 | **102**/6.2 | |
| | 4000 | | | **2.7**/1.3 | | **10.2**/1.3 | **42**/1.6 |
| | 5000 | | | | | **1.2**/0.4 | **5.2**/0.3 |
| | 6000 | | | | | | **0.8**/0.1 |

**Table 4:** Signal/background events from Herwig.

| Signal | $m_G$ [GeV] | 0.01 | 0.02 | $c=$ 0.05 | 0.07 | 0.10 | 0.20 |
|---|---|---|---|---|---|---|---|
| $e^+e^-$ | 1000 | **1128**/71 | **4543**/71 | **26640**/79 | | **91564**/160 | |
| | 1500 | **772**/12.6 | | **3313**/14.0 | | **11174**/30 | |
| | 2000 | **22**/3.5 | | **521**/3.9 | | **1958**/8.1 | |
| | 3000 | **1.3**/0.4 | | **32**/0.4 | **56**/0.5 | **115**/0.8 | |
| | 4000 | | | **2.8**/0.05 | | **10.3**/0.1 | **40.9**/0.4 |
| | 5000 | | | | | **1.2**/0.01 | **4.4**/0.1 |
| | 6000 | | | | | | **0.5**/0.01 |
| $\mu^+\mu^-$ | 1000 | **1095**/590 | **4747**/590 | **28253**/591 | | **108123**/610 | |
| | 1500 | **774**/151 | | **3213**/151 | | **12809**/154 | |
| | 2000 | **22**/48 | | **574**/48 | | **2178**/49 | |
| | 3000 | **1.3**/7.3 | | **34**/7.3 | **66**/7.3 | **134**/7.4 | |
| | 4000 | | | **3.0**/1.4 | | **12.0**/1.5 | **47.3**/1.9 |
| | 5000 | | | | | **1.3**/0.3 | **5.1**/0.4 |
| | 6000 | | | | | | **0.6**/0.1 |

The above results had to be interpolated, in order to allow plotting the 95% excluded region in ($c$, $m_G$) parameter space. Background was described by

$$N_B(c, m_G) = (A \cdot c^2 + B) \cdot e^{(C \cdot m_G^2 + D \cdot m_G + E)} \tag{11}$$

and signal by

$$N_S(c, m_G) = (F \cdot c^2 + G \cdot c) \cdot e^{(H \cdot m_G^2 + I \cdot m_G + J)} \tag{12}$$

The functions' parameters were tuned with the $\chi^2$ method to reproduce simulated event numbers. The values obtained are shown in Tab. 5.

**Table 5:** Signal and background parametrization.

| | PYTHIA | | HERWIG | |
|---|---|---|---|---|
| Signal | Electrons | Muons | Electrons | Muons |
| F | 129.1 | 79.7 | 130.1 | 86.2 |
| G | 0.36 | 0.15 | 0.34 | 0.09 |
| H | $3.04 \times 10^{-7}$ | $3.27 \times 10^{-7}$ | $3.14 \times 10^{-7}$ | $3.12 \times 10^{-7}$ |
| I | -0.0046 | -0.0047 | -0.0047 | -0.0047 |
| J | 14.57 | 15.59 | 15.57 | 16.05 |
| $\chi^2$ / NDF | 2.30 / 3 | 3.08 / 3 | 2.91 / 3 | 3.17 / 3 |
| Background | | | | |
| A | 1.42 | 0.49 | 1.43 | 0.50 |
| B | 102.9 | 187.8 | 103.2 | 187.1 |
| C | $2.54 \times 10^{-7}$ | $2.46 \times 10^{-7}$ | $3.09 \times 10^{-7}$ | $2.49 \times 10^{-7}$ |
| D | -0.0038 | -0.0033 | -0.0041 | -0.0033 |
| E | 12.18 | 10.99 | 12.49 | 11.08 |
| $\chi^2$ / NDF | 1.95 / 14 | 1.29 / 14 | 2.77 / 14 | 1.96 / 14 |

# 4 Results

The search reach of the CMS experiment is shown in Fig. 12 for Pythia, and in Fig. 13 for Herwig. The plots show 95% confidence level exclusion limits for electrons and muons. The results show that LHC should probe the whole theoretically allowed region. The contour from Herwig reaches masses ~100-200 GeV higher than Pythia, because of a higher number of selected events.

For masses under 4 TeV, the electron channel gives a higher search reach. Because of better energy resolution for electrons, the observed resonance width is smaller, resulting in a smaller mass window and a better ratio of signal to background. For higher graviton masses the background is negligible, and the effect vanishes. Because the signal in Pythia is higher for muons, this channel has a higher reach for large masses.

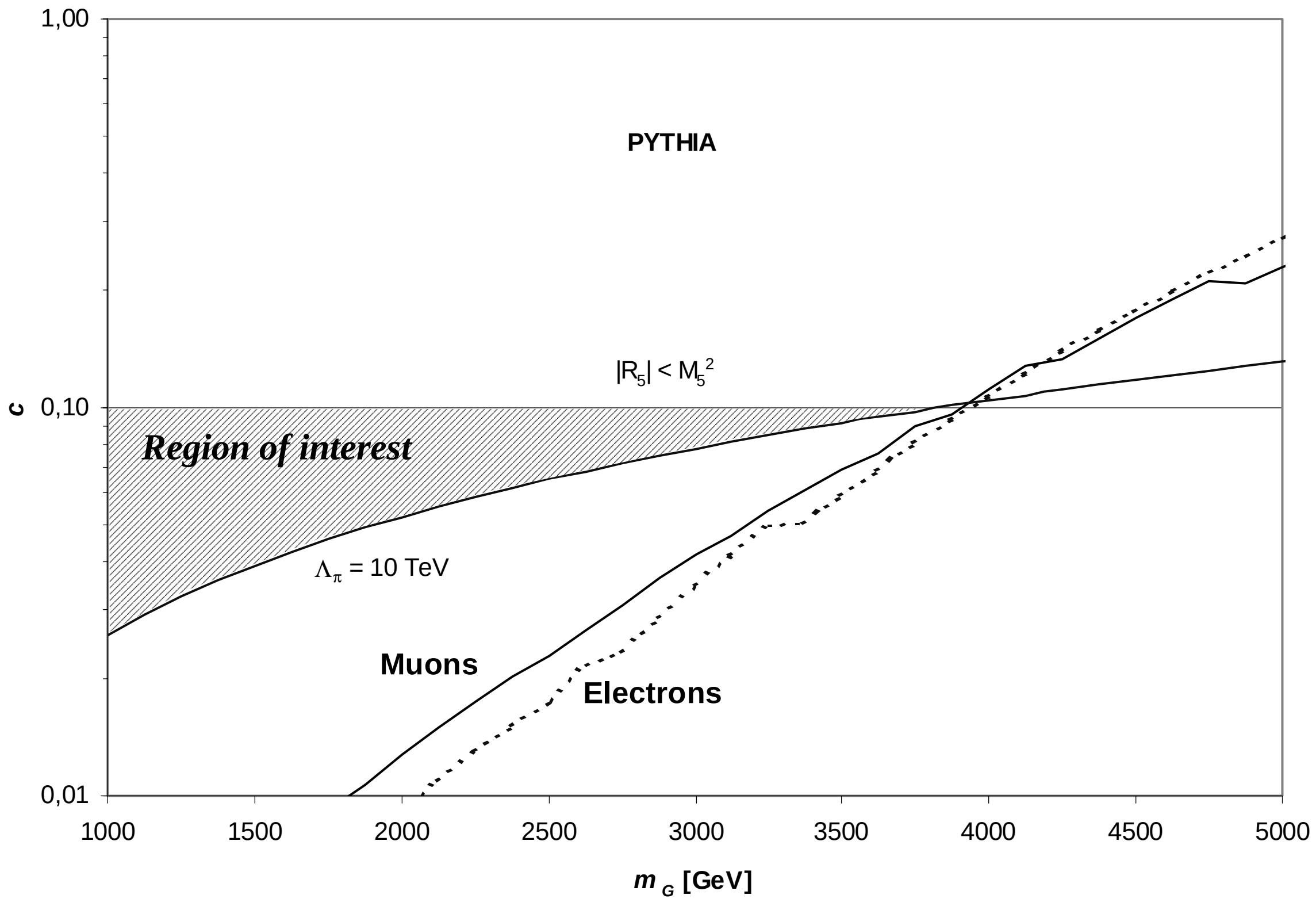

**Figure 12:** Search reach of the CMS experiment, based on results from Pythia. The 95% exclusion region lies above the curves.

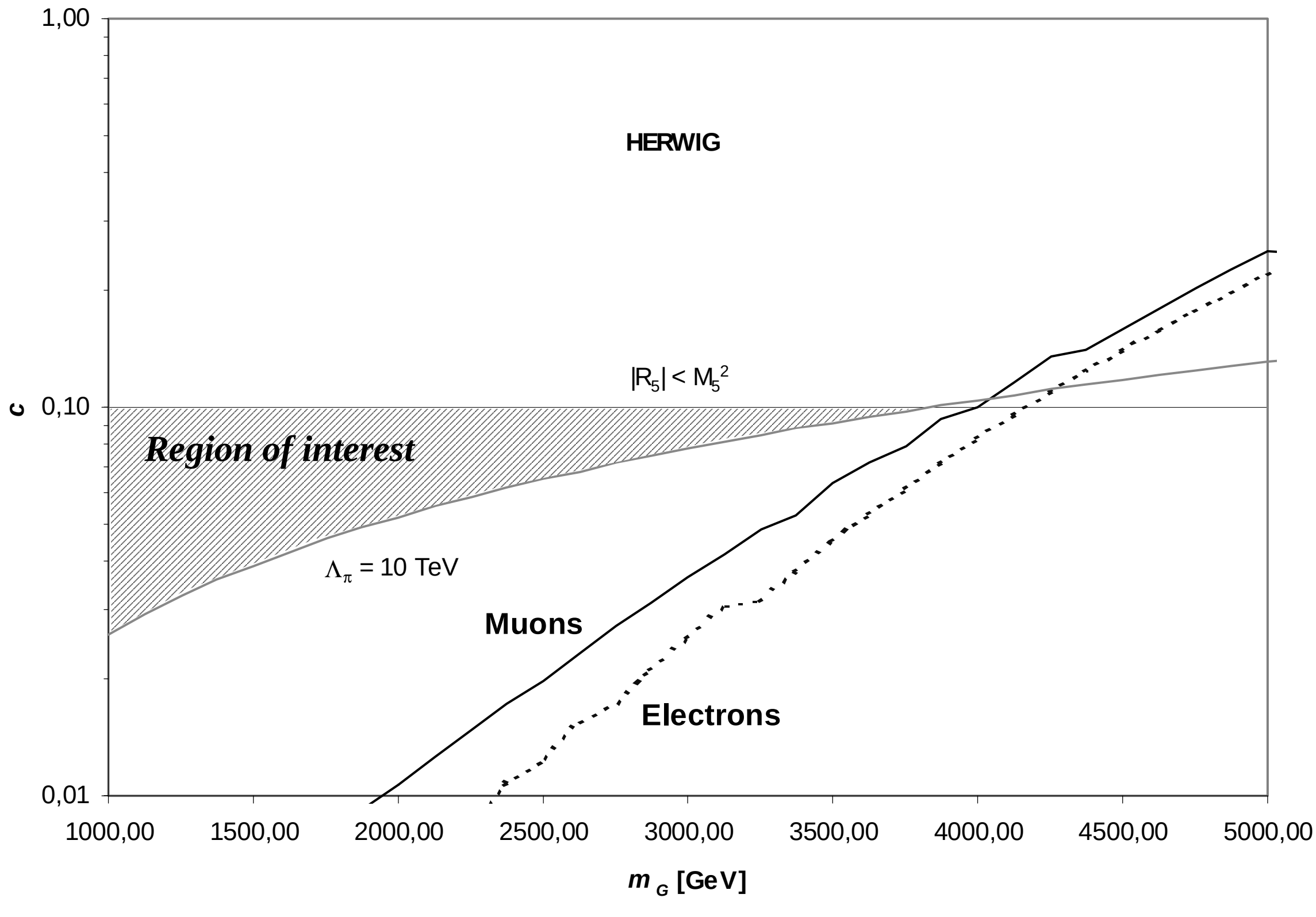

**Figure 13:** Search reach based on results from Herwig.

## 5 Summary and outlook

The analysis presented in this paper shows that the CMS experiment should be able to test the Randall-Sundrum model at 95% confidence level, after a year of CMS data taking at the design luminosity. Still, the graviton has to be distinguished from other exotic heavy particles, which requires analyzing angular distributions of the decay products. Such an analysis for the ATLAS detector is presented in [9]. According to the authors, spin-2 hypothesis is favored over spin-1 at 90% confidence level for masses up to 1440 GeV (with $c$ = 0.01). The mass reach for the $pp \to G_{KK} \to e^+e^-$ channel is 1830 GeV.